\newcommand{\pld}{\mbox {$ (\frac{2\pi}{\Lambda})^{d}$}}
\newcommand{\inl}
{\mbox {$ \int_{-\frac{\Lambda}{2}}
^{+\frac{\Lambda}{2}}\frac{d^{d}p}{(2\pi )^{d}} $}}

\documentstyle[epsf,12pt]{article}

\newcommand{\be}{\begin{equation}}
\newcommand{\br}{\begin{eqnarray}}
\newcommand{\ee}{\end{equation}}
\newcommand{\er}{\end{eqnarray}}

\newcommand{\p}{\mbox {$ \partial$}}

\begin{document}
\title{
\hfill\parbox{4cm}{\normalsize IMSC/97/03/10}\\
\vspace{2cm}
Fundamental Strings and D-strings in the IIB Matrix Model}
\author{B. Sathiapalan\\
{\em Institute of Mathematical Sciences}\\
{\em Taramani}\\
{\em Chennai 600113}\\
{\em INDIA }}
\maketitle
\begin{abstract}
The matrix model for IIB Superstring proposed by
Ishibashi, Kawai, Kitazawa and Tsuchiya is investigated.  
Consideration of
planar and non-planar diagrams suggests that 
the large N perturbative expansion is consistent with
 the double
scaling limit proposed by the above authors. We write down a Wilson
loop that can be interpreted as a fundamental string vertex operator. 
The one point tadpole in the presence of a D-string has the right form
and this can be viewed as a matrix model derivation of the boundary
conditions that define a D-string. We also argue that if world sheet
coordinates $\sigma $ and $\tau$ are introduced for the fundamental
string, then the conjugate variable $\frac{d}{d\sigma}$ and $\frac
{d}{d\tau}$ can be interpreted as the D-string world sheet
coordinates.
In this way the $SL(2Z)$ duality group of the IIB superstring becomes
identified with the symplectic group acting on ($p,q$).
\end{abstract}
\newpage
\section{Introduction}

A large N supersymmetric Yang-Mills (SSYM)  matrix model in zero space and
one time has been proposed as a non-perturbative definition of
M-theory
in the light cone gauge \cite{BFSS}.  Motivated by this proposal,
Ishibashi, Kawai, Kitazawa and Tsuchiya (IKKT) have proposed
a reduced (zero space and no time) large N SSYM  as
a non perturbative definition of type IIB string theory \cite{IKKT}.
While it is not yet clear whether there is any computational
advantage,
conceptually the matrix model approach seems very rich and fertile.
In particular a host of non perturbative phenomena involving
p-branes and D-p-branes can be incorporated into the formalism
without much difficulty.  Fundamental strings do not have a preferred
status in these models (indeed they are hard to find), which is a viewpoint that has
recently become popular.  Thus these propoals merit serious study.
Moreover,
while the idea that large N Yang Mills theory is a string theory,
is an old one, the proposals in \cite{BFSS,IKKT} are probably the
most precise and concrete that have so far been made.  Recently
several papers have appeared investigating the IIA model
\cite{Banks,Motl,Ver,Ima} and the IIB model \cite{Mak1,Mak2,Yon,F,Li}.

In this paper we focus our attention on the type IIB theory of
\cite{IKKT} and present some results that provide further evidence
for their proposal.  Taking their proposal that the Wislon loop
represents the fundamental string, at face value (this is consistent
with the old ideas of Yang Mills theory being a string theory), we
study the genus expansion of the reduced model using the ideas
of quenched Eguchi-Kawai models.  We find that a double scaling
limit very similar to that proposed by IKKT, does, in fact, define a
sensible
expansion.  We then introduce string world sheet coordinates $\sigma$
and $\tau$ to recover the standard string theory world sheet action
(in an approximation that neglects the quartic terms) in light
cone gauge, and we are also able to show that appropriately constructed
Wilson loops behave like vertex operators.  In addition, we find that
there is a symmetry between ($\sigma ,\tau $) and the conjugate
variables ($\p _{\sigma},\p _{\tau}$).  Since one set is as good
as the other, an obvious interpretation is that the conjugate
variables represent the coordinates of the D-string.  This is
also consistent with the idea that a solution of the equations of
motion represent D-strings and that fluctuations of the D-string 
solution represent fundamental strings.  This is analogous to the
fact that in field theories, classical solutions represent solitons
and the field quanta are then small fluctuations about the classical
solution.  But now, because of the complete symmetry between ($\sigma ,\tau
$)
and ($\p _{\sigma} ,\p _{\tau}$) the string and the D-string are on an
equal
footing and the $SL(2,Z)$ symmetry of type IIB is then nothing more
than the symplectic group acting on $(p,q)$i.e.
\[
\left( \begin{array}{c} p' \\ q' \end{array} \right) =
\left( \begin{array}{cc} a & b \\ c & d \end{array} \right)
\left( \begin{array}{c} p \\ q \end{array}\right)
\]
We then look at the one point function of the Wilson loop in the
presence
of the D-string and check that it reproduces the result of
\cite{IKKT}.
In the usual formalism, this result is a direct consequence of the
boundary
conditions on the open string that define a D-string \cite{Polch}.  Thus we
can view this calculation as a matrix model derivation of the
boundary conditions.

This paper is organized as follows.  In section II we discuss the
diagrammatic expansion of the reduced model and check the double
scaling proposal.  In section III we discuss in some detail the
introduction of the world sheet coordinates for the fundamental
(`F')string and the D-string and argue that the symplectic group
should
be identified with the SL(2,Z) duality of IIB strings.  We also
study the spectrum and identify the (1,m) strings.  In section IV
we study some properties of the Wilson loop, namely the two point
function in flat space and the one point function in the presence
of a classical D-string background.  We conclude with some comments in
Section V.

\newpage
\section{Perturbation Theory with the Reduced Model}

Reduced models \cite{EK,BHN,Par,DW,GK} were originally proposed as a large N
approximation to the full theory, the main point being that the
planar diagrams give the same result to leading order in $\frac{1}{N}$
as in the full theory.  In the present context the reduced model at
infinite $N$ \footnote{$N$ needs to be literally infinite if one is to get
anything non trivial that looks like
the D-string of the continuum theory in some non zero space time
dimension.} is being taken as the defining theory.
  Therefore, unlike
in the early days, it makes sense to consider the subleading non-planar
diagrams of the theory and check whether there is some kind of
double scaling limit that can make sense.  In the original context
the perturbation series in Yang-Mills is of the form 
$(g^{2}N)^{\frac{V_{3}}{2}+V_{4}}(N^{2})^{1-g}$. Here
$V_{3}$ and $V_{4}$ are the numbers of cubic and quartic
vertices respectively, and $g$ is the genus.  The genus
expansion is labelled by powers
of $\frac{1}{N}$.  Clearly in the reduced model, since we need to
take $N \rightarrow \infty$ and not just large, this would not work.
Thus there must be some other double scaling at work here.
The idea that the reduced model
perturbation could define a string in some double scaling limit has
been proposed earlier \cite{Bars}.

In \cite{IKKT} a double scaling limit has been proposed, based on
comparison with the classical action of the D-string.  Below we
consider the perturbation expansion of this model.  We find that
the double scaling limit proposed in \cite{IKKT} is the correct
one.  We follow the prescription of \cite{DW,GK}.  Since
we are only doing some simple counting and are not going to actually
evaluate any diagrams, the details of the vertices are not important.
The action of the SSYM is
\be     \label{2.1}
S \; = \; \int d^{10}x\frac{1}{2g^{2}}[F_{\mu \nu}]^{2} + \; Fermionic
\ee
Note that if we assign the usual canonical dimension of one to $A$
(in mass units), then $g$ has dimensions of $[mass]^{-3}$ or
$[length]^{3}$.  Thus we can write 
\be     \label{2.2}
g^{2} = \frac{g_{0}^{2}}{N}
=\frac{\lambda ^{2}}{N}[m]^{-6}
\ee
We have factored out $\frac{1}{N}$, in anticipation of the fact that
($g^{2}N$) is usually held fixed.  Here ``$m$'' is some finite mass
scale
the theory is assumed to possess.  In the reduced model we write
\be     \label{2.3}
S_{Reduced}= \pld \frac{1}{2g^{2}}[A_{\mu},A_{\nu}]^{2} + \; Fermionic
\ee
following \cite{DW,GK}. $[\frac{2 \pi}{\Lambda}]^{d}$ can be taken as
the volume of a basic cell, where $\Lambda$ is a momentum cutoff. Actually
in this model, since the eigenvalues of $A$ represent position, 
$\Lambda$ should be interpreted as an infrared cutoff in the real
space-time sense.  Neverthelss, the mathematical manipulations
are the same.

In order to do perturbation theory we assume that
\be     \label{2.4}
A_{\mu} \; = \; P_{\mu} + a_{\mu}
\ee
``$P_{\mu}$'' can be thought of as the vev of $A_{\mu}$ and this 
becomes equivalent to the usual prescription for introducing a
``momentum" dependence that is associated with the group index.  
\footnote{As mentioned above, the $A_{\mu}$ have the interpretation of
position of the D-branes. In the context of the quenched reduced  
models \cite{EK,BHN,Par,DW,GK} however, this would correspond to momentum.}

We take
$P_{\mu}$ to be diagonal
\be     \label{2.5}
P_{\mu} = \left( \begin{array}{cccc}
p_{\mu}^{1} &0           &      &         \\  
0           & p_{\mu}^{2}&      &         \\  
            &            & ...  &         \\
            &            &      & p_{\mu}^{N}
\end{array} \right) 
\ee
where $P_{\mu}^{i}$ belong to the set of eigenvalues of momentum
in the space-time of interest, and assumed to be uniformly
distributed. 
Then we recover the YM action in the quenched momentum prescription,
with a kinetic term and cubic and quartic interaction vertices.
\be     \label{2.6}
S = \pld \frac{1}{2g^{2}}\{ [P_{\mu},a_{\nu}]^{2} + 2[P_{\mu},a_{\nu}]
[a_{\mu},a_{\nu}] + [a_{\mu},a_{\nu}]^{2}\}
\ee 
We also have to include gauge fixing terms as in \cite{DW,GK}
but for our purposes the above action will suffice.
Finally we can use
\be     \label{2.7}
\frac{1}{N}\pld \sum _{i=1}^{N} f(p_{\mu}^{i}) \rightarrow \inl 
f(p_{\mu}^{i})
\ee
We also have the following relations:
\[
P=V_{3}+V_{4}+L-1      
\]
\be     \label{2.8}
2P=3V_{3} +4V_{4}
\ee
where $P,V_{3},V_{4},L$ are the number of propagators, cubic and
quartic vertices and loops respectively.  Thus keeping in mind the
following
planar diagram (Figure 1) for concreteness, and using the fact that there
are factors of $g\pld $ associated with a cubic vertex, $g^{2}
\pld $ with a quartic vertex and $\frac{1}{(p_{i}-p_{j})^{2}\pld }$
with a propagator, we get
\be     \label{2.9}
(g^{2})^{\frac{V_{3}}{2} +V_{4}}[\pld ]^{V_{3}+V_{4}-P}
\underbrace{\sum _{i,j...}}_{L+1}
f(p^{i},p^{j},...)
\ee
\begin{center}
\leavevmode
\epsfxsize=3in \epsfbox{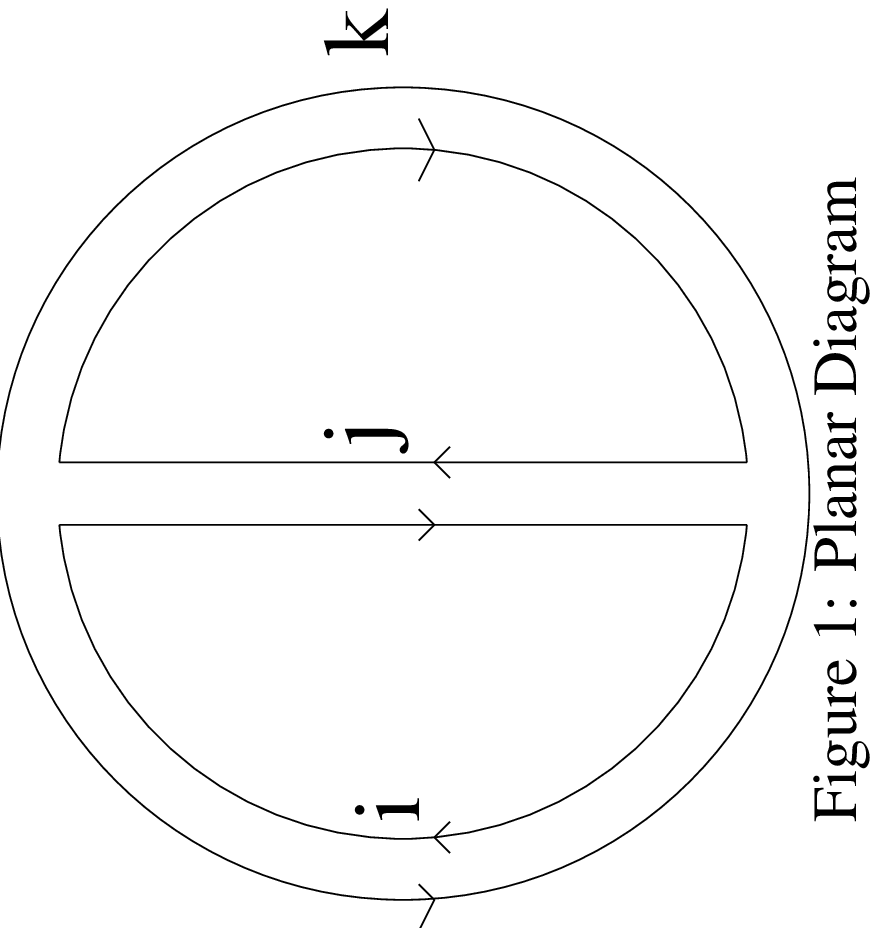}
\end{center}
Now we replace $\Sigma _{i}$ by $\pld N \inl $.  Since only
differences of the form $p_{i}-p_{j}$ are involved, there is
one overall factor of $N$, not accompanied by any momentum
integration.  This gives
\be     \label{2.10}
N^{L+1} [\pld ]^{L} (g^{2})^{L-1}[\pld ]^{(1-L)} \underbrace{\inl ...} 
_{L}f(p_{i}...)
\ee
\be     \label{2.11}
= \; N^{2} (g^{2}N)^{L-1} \pld [Usual \; Feynman \; diagram \; +\; O(1/N)]
\ee
By usual Feynman diagram, we mean that of 
the unreduced model.  The $\frac{1}
{N}$ corrections are due to the fact that in some of the terms in the 
sum (\ref{2.9}), the propagator may have zero momentum, being of the
form
``$p_{i}-p_{j}$'' with $i=j$. In our case, since we are not concerned
with separating out a piece relevant to the unreduced model, the full
expression in square brackets is the answer.  The overall volume
factor of $\pld$ is expected and one divides it out to get the free
energy per unit volume.

We can now consider a non planar diagram with one handle as in 
Figure 2. 
\begin{center}
\leavevmode
\epsfxsize=3in \epsfbox{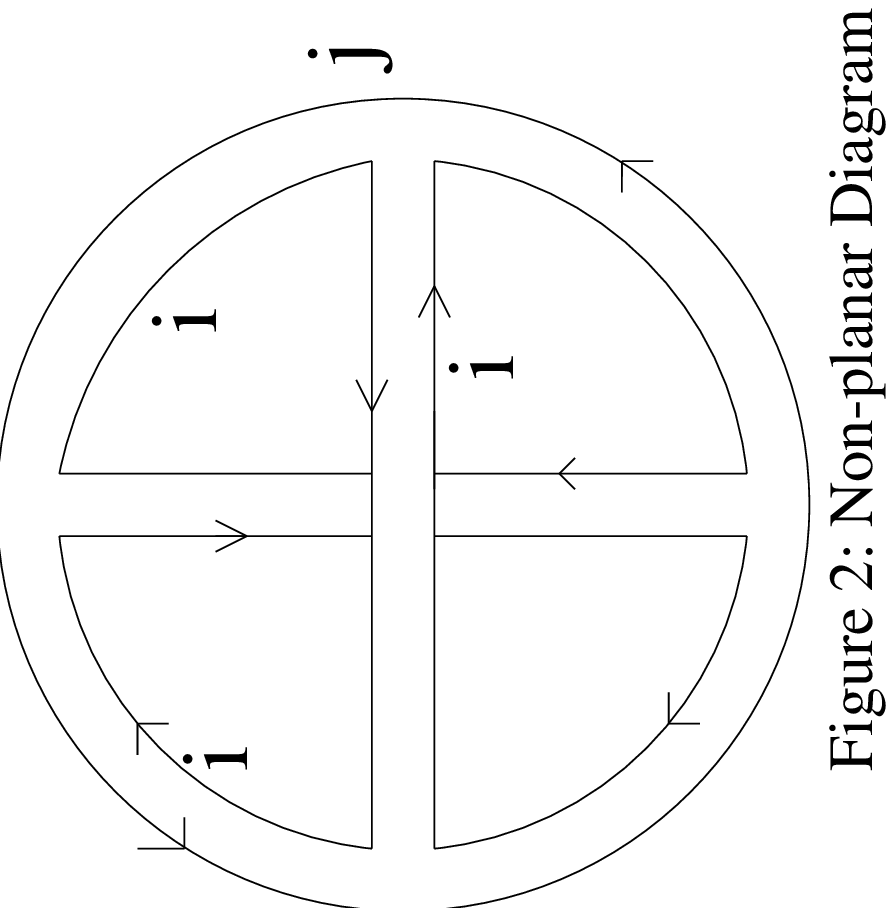}
\end{center}

 Note that
while there are three loops in the usual sense
 there are only two index loops.  Thus we get a $N^{L-1}$ rather than
$N^{L+1}$.  In general we get $N^{L+1-2H}$ where $H$ is the number
of handles.  Since momentum integration is associated with the group
index, there are only two of them :$\sum _{i}\sum_{j}$ and again
since only $p_{i}-p_{j}$ is involved, we effectively have one
integration.
Thus $\sum _{i}\sum _{j} f(p_{i},p_{j}) = N^{2}
\pld \inl f(p)$, for this particular example.  In general
it is $N^{L+1-2H}[\pld ]^{L-2H}\underbrace{[\inl ...}
_{L-2H \; of \; them}]f(p)$.

Thus for a non planar diagram (\ref{2.11}) gets replaced by
\be     \label{2.12}
N^{2}(g^{2}N)^{L-1} [N \pld ]^{-2H} [Momentum \; integral]
\ee
The expression in square brackets has fewer momenta integrated
over and is therefore more convergent than that of the leading term
(\ref{2.11}).  If we assume that these integrals are convergent for 
reasons of supersymetry and that the perturbation makes sense, then we
see clearly that the genus expansion parameter is not 
$\frac{1}{N}$ 
but 
$\frac{1}{N \pld }$.  
Thus we can take 
$N$ to 
$\infty$, along
with 
$\Lambda ^{d}$ (keeping $g^{2}N$ fixed as usual) 
and get a non zero contribution from the non
planar diagrams.
We can thus identify the string coupling constant $g_{s}$
\be     \label{2.13}
g_{s}=\frac{1}{N\pld m^{d}}
\ee
where ``$m$'' is some finite mass scale characterizing the theory. 
This is essentially the double scaling limit advocated in \cite{IKKT},
with $d \; = \; 10$.

However, there is an ambiguity here.
We could also imagine starting with the same model in a lower spacetime
dimension.  This corresponds to giving an expectation value to only
some of the $A_{\mu}$.  This should then correspond to a reordering
of the perturbation series where the propagator is expanded out, viz:
\be     \label{2.14}
\frac{1}{\sum_{\mu =1}^{10}p^{\mu}p_{\mu}} \; = \; 
\frac{1}{\sum_{i =1}^{d}p^{i}p_{i}}[1+ \frac{\sum _{j=d+1}^{10}p^{j}p_{j}}
{\sum_{i=1}^{d}p^{i}p_{i}} + \; ... ]
\ee
The numerators in the above series correspond to the effects
of quartic interaction vertices where two of the $A_{\mu}$ take 
vevs $<A_{i}>=p_{i}$.  
In particular one could start with $d=2$ and in that case we would
interpret it as the string world sheet.  The full space time
interpretation would emerge on calculating string diagrams
with sources (``vertex operators'') that have the effect
of the non leading terms in (\ref{2.14}). 
Presumably, since we are only reordering perturbation theory, one
should still have the original double scaling with $\frac{N}{\Lambda ^{10}}$.
  But
in each individual graph this would not make sense - one would need
${N}{\pld}$.  One possible explanation is that the double scaling
should
be chosen according to the vacuum that one is expanding about and
different
limits correspond to exploring different vacua. 
We have to also keep in mind
that the space-time interpretation in string theory requires the
``$p$'' to be position coordinates.  In the Yang Mills description
 they are momenta.

To conclude this section, if we accept the conjecture that the $d=10$ SSYM defines
string theory and that perturbation about flat 10-d space time makes 
sense, then the double scaling prescription (\ref{2.13}) of IKKT seems
to be the right one.

\newpage
\section{World Sheet Coordinates for the Fundamental String and
the D-string}
\setcounter{equation}{0}

If we accept that the large $N$ reduced SSYM action of \cite{IKKT}
describes the IIB theory we can write the action as
\be     \label{3.1}
S \; = \; \frac{Tr}{\alpha ^{'2} g_{s}}\{[A^{\mu},A^{\nu}]^{2} +
\bar{\Psi} \Gamma ^{\mu} A_{\mu} \Psi \} 
\ee
Noting that the 10-dimensional gravitational constant $\kappa \approx
\alpha  ^{'2} g_{s}$ we can rewrite (\ref{3.1}) as 
\be     \label{3.2}
S = \frac{1}{\kappa }Tr[A^{\mu},A^{\nu}]^{2} + \; Fermionic
\ee
One advantage of introducing $\kappa$ in place of $\alpha ^{'2}g_{s}$
is that it looks more duality invariant - it does not give a special
role to $\alpha '$.  This is particularly important for type IIB
where there is a complete duality between F-strings and D-strings.
While this is a non perturbative result, we should expect to see this 
symmetry in the matrix model formulation which purports to being a 
non perturbative definition of string theory.  While keeping this
requirement
in mind, following the suggestion of IKKT that the Wilson loop
should correspond to the F string, let us introduce
world sheet coordinates $\tau _{F}$ and $\sigma _{F}$
for the F-string by giving expectation values to $A^{0}$
and $A^{1}$.  This forces us into the light cone gauge.
\footnote{We have not discussed the fermion sector in this paper.
This will be reported elsewhere \cite{TJ}.}  

We use the fact that a general $N \times N$ matrix can be expanded in a
basis.
\be     \label{3.3}
A = \sum _{m,n} A_{mn}e^{imp}e^{inq}
\ee
where
\be     \label{3.4}
[q,p] = \frac{2\pi i}{N}
\ee                                 
There is the usual caveat that $N$ has to be infinity for this
to work.  To make things concrete, consider a particle in a one
dimensional box of length $L$, with periodic boundary conditions.
Also assume an ultraviolet cutoff i.e. a lattice spacing ``$a$''.
The number of sites is $L/a \equiv N$.
We introduce canonical momentum and position operators $P$ and $Q$.
The momentum eigenfunctions are of the form
\be     \label{3.6}
\phi _{n} \; = \; \frac{1}{\sqrt{L}}e^{ip_{n} x}
\ee
where $p_{n}=\frac{2\pi n}{L} =\frac{2\pi n}{N}a$ is the $n ^{th}$ eigenvalue
of the momentum operator $P$.  Let us choose
$a = \sqrt {\frac{2 \pi}{N}}$ and $L=\sqrt {N\pi}$.  Then
\be     \label{3.61}
p_{n} = \frac{2\pi n}{\sqrt{N2\pi}}  \: \: \: 0\leq n\leq N-1
\ee
The operator $p=P \sqrt{\frac{2\pi}{N}}$, has eigenvalues 
$\frac{2\pi n}{N}$. The matrix $e^{iP \sqrt{\frac{2\pi}{N}}}$ then is
\be     \label{3.7}
e^{ip} = e^{iP \sqrt{\frac{2\pi}{N}}} =
\left( \begin{array}{ccccc}1 & 0      &   &    &              \\ 
                           0 & \omega &   &    &              \\ 
                             &        &   & 0  &              \\
                             &        & 0 & ...&              \\
                             &        &   &    & \omega ^{N-1}\\  
                            \end{array} \right)
\ee
where $\omega$ are the $N^{th}$ roots of unity.
$Q$ has possible values $0,a,2a,...,na,..(N-1)a$ which is the same
as
\be     \label{3.8}
(0,\frac{1}{\sqrt{N}},.....,\frac{N-1}{\sqrt{N}})\sqrt{2\pi}
\ee
the same as $ P$.  In this normalization $[Q,P]=i$.
If I choose $q=\sqrt{\frac{2 \pi}{N}}Q$ so that
\be     \label{3.10}
[q,p]=\frac{2 \pi i}{N}
\ee 
it has eigenvalues $\frac{n}{N}2\pi$, the same as $ p$.
\be     \label{3.11}
e^{iq}=\left( \begin{array}{ccccc}0 & 1 &  &  &  \\
                            0 & 0 & 1 &  & \\  
                              &   & 0 & 1 & \\
                              &   &   & .. & 1\\
                            1 &   &   &    & 0 \end{array} \right)
\ee
Note again that the matrix $q$ satisfying (\ref{3.10}) and (\ref{3.11})
only exist in the limit $N \rightarrow \infty$ which is also
the limit $L \rightarrow \infty $ or $ a \rightarrow 0$.

We introduce $\tau _{F}$ and $\sigma _{F}$ by the following
ansatz
\be     \label{3.12}
[A^{0}]^{jb}_{ia} \; = \; \frac{1}{\sqrt{2}} \kappa ^{1/4}g_{s}^{1/4}
[-i\frac{\partial}{ \partial \tau _{F}}]^{j}_{i} \delta _{a}^{b} +
 [a^{0}]_{ia}^{jb}
\ee
\be     \label{3.13}
[A^{1}]^{jb}_{ia} \; = \; \frac{1}{\sqrt{2}} \kappa ^{1/4}g_{s}^{1/4}
[-i\frac{\partial}{\partial \sigma _{F}}]^{b}_{a} \delta _{i}^{j} +
 [a^{1}]_{ia}^{jb}
\ee
$i,j=1.....N_{1}$ $a,b=1,   N_{2}$. 
We can set
\[
2\pi \tau _{F} =q_{1} \: ;\: \: -i\frac{\partial}{\partial \tau _{F}}
=
 N_{1}p_{1}
\]
\be     \label{3.14}
2\pi \sigma _{F} =q_{2} \: ; \: \: -i\frac{\partial}{\partial \sigma
_{F}} =
 N_{2} p_{2}
\ee

One can, if one wants, let $2\pi \tau _{F} = N_{1}q_{1}$ and
let $-i \p _{\tau _{F}} = p_{1}$ and remove the factor
$N_{1}$ in (\ref{3.16}).  This may be more convenient
for some purposes.  However, in this paper we have treated 
$\sigma $ and $\tau$
symmetrically.

Here $a^{0}$ and $a^{1}$ are fluctuations about the assumed background.
Thus
\be     \label{3.16}
<A^{0}> = \frac{1}{\sqrt{2}}\kappa ^{1/4} g_{s}^{1/4}[p_{1}]_{i}^{j}
\delta _{a}^{b} N_{1}
\ee
\be     \label{3.17}
<A^{1}> = \frac{1}{\sqrt{2}}\kappa ^{1/4} g_{s}^{1/4}[p_{2}]_{a}^{b}
\delta _{i}^{j} N_{2}
\ee
The action (\ref{3.2}) becomes [$N_{1}N_{2}=N$]
\be     \label{3.18}
S=\frac{\sqrt{g_{s}}}{2\sqrt{\kappa}}[(F_{01})^{2} \; + \;
[D_{\tau F}A^{I}]^{2} + [D_{\sigma F}A^{I}]^{2}] 
+\frac{1}{\kappa}\sum _{I<J}[A^{I},A^{J}]^{2}
\ee
Here $D_{\tau F}$ and $D_{\sigma F}$ are covariant derivatives
in the $\tau _{F}$ and $\sigma _{F}$ directions. 
If we ignore the interaction terms (and $F_{01}$) and assume that
$A^{I}$ is only a function of $\tau _{F}$  and $\sigma _{F}$,
then they are like the transverse string coordinates.  This is
the light cone gauge string action. We will
also work out, in the next section, the two point function for a Wilson loop
and show that the Wilson loop is nothing but a vertex operator.

If we look at (\ref{3.18}) it also looks like the action that describes the
fluctuations of a D-string, except for the factor 
$\sqrt{\frac{g_{s}}{\kappa}}$ (=$\frac{1}{\alpha '}$) which reminds us
that this is supposed to be a fundamental string action.  Before we
worry about this source of confusion, let us consider the classical
D-string solution of \cite{IKKT}.  In our language it corresponds to
turning on an electric field by giving an expectation value
to $a^{0}$ and $a^{1}$ in (\ref{3.12}) and (\ref{3.13})\footnote{
Similar observations have been made in \cite{Ver}.Some rescalings
have to be done to make precise contact}.  Thus we let
\be     \label{3.19}
<a_{0}> = 
\frac{1}{\sqrt{2}}\kappa ^{1/4} g_{s}^{-1/2}\frac{\sigma _{Fa}
^{b}\delta _{i}^{j}}{(N_{1}N_{2})^{1/4}}
=\; \frac{1}{\sqrt{2}}\kappa ^{1/4} g_{s}^{-1/2}\frac{q_{2a}
^{b}\delta _{i}^{j}}{2\pi(N_{1}N_{2})^{1/4}}
\ee
\be     \label{3.20}
<a_{1}> = -
\frac{1}{\sqrt{2}}\kappa ^{1/4} g_{s}^{-1/2}\frac{\tau _{Fi}
^{j}\delta _{a}^{b}}{(N_{1}N_{2})^{1/4}}
=\; -\frac{1}{\sqrt{2}}\kappa ^{1/4} g_{s}^{-1/2}\frac{q_{1i}
^{j}\delta _{a}^{b}}{2\pi(N_{1}N_{2})^{1/4}}
\ee
Thus $<A^{0}>_{total}$ is now the sum of (\ref{3.16}) and (\ref{3.19}) and
similarly for $<A^{1}>_{total}$. $<A^{0}>_{total}$ and
$<A^{1}>_{total}$
do not commute now and
\be     \label{3.21}
[<A^{0}>_{total},<A^{1}>_{total}]=
F_{01} = \;i \frac{\sqrt{\kappa}g_{s}^{-1/4} 
\delta _{i}^{j} \delta _{a}^{b}}{(N_{1}N_{2})^{1/4}}
\ee

\be     \label{3.22}
S_{classical} \; =\; 
\frac{Tr}{\kappa}(F_{01})^{2} \; = \;
\frac{\sqrt{N_{1}N_{2}}}{\sqrt{g_{s}}} = \;
\frac{TL}{\sqrt{\kappa g_{s}}}
\ee
We have set $\sqrt{N_{1}} = T$ and $\sqrt{N_{2}} =L$, both
in units of $\kappa ^{1/4}$. (Note that $Q_{1,2}$, which is the 
one that has canonical commutation relations,
has a range of
$\sqrt{2\pi N_{1,2}}$ whereas the world sheet coordinate
$q_{1,2}$ has a range of $2\pi$.)
This gives a tension of O($\frac{1}{\sqrt{g_{s}}}$)in Planck units, as
required for D-strings.  In this picture fluctuations 
(that depend on $\tau_{F},\sigma_{F}$) about this
classical solution are fundamental strings.

Now we notice something interesting. $p$ and $q$ have the
same spectrum. We could have let 
\be     \label{3.34}
<A^{0,1}> =
 \frac{1}{\sqrt{2}}(\frac{\kappa}{g_{s}})^{1/4}q_{1,2}N_{1,2}
\ee
instead of $p_{1,2}$ as in (\ref{3.16}),(\ref{3.17}).  The entire analysis
would have gone through in exactly the same way.  We can introduce
worldsheet coordinates $\tau _{D}$ and $\sigma_{D}$ and set 
\be     \label{3.35}
N_{1}q_{1} = -i\frac{\partial}{\partial \tau _{D}} \; ;\; \; p_{1} = 
2 \pi \tau _{D}
\ee
\be     \label{3.36}
N_{2}q_{2} = -i \frac{\partial}{\partial \sigma _{D}}\; ; \; \; p_{2}
= 
2 \pi \sigma
_{D}
\ee
 We get an action identical to (\ref{3.18}) with the subscript
F replaced by D and having an inverse string tension $\beta ' =
\alpha ' g_{s}$:
\be
S=\frac{1}{2\sqrt{\kappa g_{s}}}[(F_{01})^{2} \; + \;
[D_{\tau D}A^{I}]^{2} + [D_{\sigma D}A^{I}]^{2}] 
+\frac{1}{\kappa}\sum _{I<J}[A^{I},A^{J}]^{2}
\ee
As before, $D_{\tau D}$ and $D_{\sigma D}$ are covariant
derivatives in the $\tau_{D}$ and $\sigma _{D}$ directions.
  We interpret this as the D-string action.  We have
to impose that $A^{I}$ are functions of $\tau_{D}$ and $\sigma _{D}$
in contrast to the F-string configurations which were functions
of $\tau_{F}$ and $\sigma _{F}$.  Thus the interchange of $p
\leftrightarrow q$ and $g_{s} \leftrightarrow \frac{1}{g_{s}}$ is the
SL(2,Z) duality that interchanges the F and D-strings.  Note that
\be     \label{3.37}
\left( \begin{array}{c}p' \\ q' \end{array} \right) \; = \; 
\left( \begin{array}{cc} a & b \\ c & d \end{array} \right)
\left( \begin{array}{c} p \\ q \end{array}\right)
\ee
preserves the commutation relations, provided $ad-bc=1$.  Since 
$p$ and $q$ have identical discrete spectra, $a,b,c,d$ have to
be integers.

In Equation (\ref{3.18}) small fluctuations about a classical
solution are described by variables $a^{0,1},A^{I}$.  In general
they can be functions of $q$ and $p$, i.e. functions of
$\sigma _{F}, \tau _{F}$ as well as $\frac{\partial}{\partial
\tau_{F}},
\frac{\partial}{\partial \sigma _{F}}$,
which
is what we have been calling $\tau_{D}$ and $\sigma _{D}$.
Fluctuations that are purely functions of ($\tau _{F}, \sigma _{F}$)
or ($\tau _{D}, \sigma _{D}$) are F-strings and D-strings respectively.
If we were to substitute into the quartic interaction term, the most
general configuration, we would get terms that, viewed from the
F-string (or D-string) world sheet perspective, involve higher
derivatives in $\tau , \sigma$.  These look somewhat like massive
mode backgrounds and should play a part in determining the 
superstring interactions.

The viewpoint presented here is not specific to IIB and applies to IIA
also.  However IIB has an exact SL(2,Z) symmetry
and the above observations are more readily applicable.  Thus to
summarize,
if the above interpretation is correct, then it is very simple to
consider F and D-string configurations on the same footing at the same
time and duality is manifest.  The classical solutions break
the symmetry through particular choices of the expectation values of
the
matrices, which are chosen to reproduce either D- or F- string
backgrounds.  These solutions are distinguished only by the
relative amounts of $p$ and $q$ in the expectation value of $A$.

Given a classical D-string background (\ref{3.18}), one can
 quantize
collective fluctuations about this background.  The Wilson loop
variables $\delta a_{1}L$ are like one dimensional rotors with
quantized energies.  As pointed out in \cite{Wit} $e^{in a_{1}L}$
are eigenstates of the electric field operator.  The energy is of
O($g_{s}$) (in F-string units).  
These are thus the (n,1) strings. If we start with a multiply
wound classical background one should get the (n,m) strings.
It should not be too hard to
write everything in a manifestly SL(2,Z) covariant manner to
reproduce the spectrum of BPS states as done in \cite{JS}.

\newpage
\section{Wilson Loop and Vertex Operators}
\setcounter{equation}{0}
In (\ref{3.18}), the action for the transverse $A^{I}$ is
\be     \label{4.1}
\frac{Tr}{\kappa}\{[<A^{0}>,A^{I}]^{2} + [<A^{1}>,A^{I}]^{2} +
\sum _{I<J}[A^{I},A^{J}]^{2} \}
\ee
If we use (\ref{3.16})
\be     \label{4.2}
<A^{0}> = \frac{1}{\sqrt{2}}\kappa ^{1/4} g_{s}^{1/4}[p_{1}]_{i}^{j}
\delta _{a}^{b} N_{1}
\ee
\be     \label{4.3}
<A^{1}> = \frac{1}{\sqrt{2}}\kappa ^{1/4} g_{s}^{1/4}[p_{2}]_{a}^{b}
\delta _{i}^{j} N_{2}
\ee
It is easiest to expand $A^{I}$ as
\be     \label{4.4}
A^{I}=\sum _{m,n,r,s}A_{mnrs}e^{imp_{1}}e^{inq_{1}}e^{irp_{2}}e^{isq_{2}}
\ee
Using $[q_{1},p_{1}]= \frac{2 \pi i}{N_{1}}$
\be     \label{4.7}
[<A^{0}>,A^{I}]=\frac{\kappa ^{1/4}g_{s}^{1/4}}{\sqrt{2}}
\sum_{m,n,r,s}A_{mnrs}2\pi n e^{imp_{1}}e^{inq_{1}}e^{irp_{2}}e^{isq_{2}}
\ee
This is to be plugged into the action.
To evaluate the trace we introduce the basis vectors
$| \tilde{q}>$, with $<\tilde{q}|\tilde{q}> = 1$
and $|\tilde{p}> = \sum
_{\tilde{q}}e^{i\tilde{p}\tilde{q}}|\tilde{q}>$,
\be     \label{4.11}
<\tilde{p}| \tilde{p}>=\sum _{q}1 = N_{1}
\ee
Thus 
\[
Tr[<A^{0}>,A^{I}]^{2} \; = \; 
\]
\be     \label{4.15}  
-2\pi ^{2}\sqrt{g_{s}\kappa}N_{1}N_{2}\sum_{m,n,r,s} n^{2}
A_{mnrs}A_{-m-n-r-s}e^{2\pi i \frac{mn}{N_{1}}}e^{2\pi i \frac{rs}{N_{2}}}
\ee  
Similarly
\[
Tr[<A^{1}>,A^{I}]^{2} =
\]
\be     \label{4.16} 
-2\pi ^{2}\sqrt{g_{s}\kappa}N_{1}N_{2}\sum_{m,n,r,s}   r^{2}
A_{mnrs}A_{-m-n-r-s}e^{2\pi i \frac{mn}{N_{1}}}e^{2\pi i \frac{rs}{N_{2}}}
\ee  
The quadratic part of the action becomes
\be     \label{4.17}
-2\pi ^{2}\sqrt{\frac{g_{s}}{\kappa}}N_{1}N_{2}\sum_{m,n,r,s}(n^{2}+r^{2})
A_{mnrs}A_{-m-n-r-s}e^{2\pi i \frac{mn}{N_{1}}}e^{2\pi i \frac{rs}{N_{2}}}
\ee  
Thus 
\be     \label{4.18}
<<A_{mnrs}A_{abcd}>> = \sqrt{\frac{\kappa}{g_{s}}}\frac{\delta _{m+a}
\delta _{n+b} \delta _{r+c} \delta _{s+d} e^{-2\pi i \frac{mn}{N_{1}}}
e^{-2\pi i \frac{rs}{N_{2}}}}{2\pi ^{2}N_{1}N_{2}(r^{2}+n^{2})}
\ee  
 is the two point function.
\subsection{Two point function for the Wilson Loop}
The Wilson loop in this model can be defined as \footnote{The name
``Wilson Loop'' is a misnomer, since for generic $K$, it is
not invariant under $SU(N)$ transformations.  However
for want of a better name we will refer to it as a Wilson loop.}
\be     \label{4.19}
Tre^{KA} = \sum_{ia}[e^{KA}]_{ia}^{ia}
\ee  
where $K^{jb}_{ia}$ is a general $N_{1}N_{2}\times N_{1}N_{2}$ matrix.
Let us calculate
\be     \label{4.20}
<Tre^{KA} Tre^{\tilde{K}A}>
\ee  
Expanding each exponential and using the expansion (\ref{4.4})
for both $K$ and $A$, 
and also using 
\be     \label{4.22}
<q|e^{i(m+m')p}e^{i(n+n')q}|q> = e^{i(n+n')q}\delta _{m+m'}
\ee
we get to lowest order (suppressing the indices on $p,q,N$, so that
each of the $m,n..$ are two dimensional vectors)
\be     \label{4.23}
\sum_{q,q'}\sum_{m,n,m',n'}\sum_{r,s,r',s'}K_{m'n'}\tilde{K}_{r's'}
<<A_{mn}A_{rs}>>e^{i(n+n')q}\delta_{m+m'}e^{i(s+s')q}\delta_{r+r'}
e^{2\pi i \frac{r'm}{N}}e^{2\pi i \frac{s'r}{N}}
\ee
Let us also set 
\be     \label{4.25}
K_{m'n'}=k\delta _{m',0}\delta _{n',0}    \; \; ; \; \; 
\tilde{K}_{r's'}=\tilde{k} \delta _{r',0}\delta _{s',0}
\ee
as a special choice.  We find (\ref{4.23}) becomes
\be     \label{4.27}
k.\tilde{k}\sum_{q,q'} \sum _{n}\frac{e^{in(q-q')}}{2\pi^{2}Nn^{2}}
\ee
Restoring the two dimensional nature we see that we have in the
continuum limit:
\be     \label{4.28}
k.\tilde{k}\int d^{2}q \int d^{2}q' \frac{ln(q-q')}{N}
\ee
At the next order we get
\be     \label{4.29}
(k.\tilde{k})^{2} \int d^{2}q \int d^{2}q' \frac{\ln^{2}(q-q')}{2N}
\ee
Thus it is easy to see that it exponentiates to give:
\be     \label{4.30}
<Tr e^{KA} Tr e^{\tilde{K}A}> = \int d^{2}q \int d^{2}q' e^{k.\tilde{k}
ln(q-q')}
\ee

Thus we see that $Tr e^{KA}$ behaves like $e^{ikX}$ in ordinary
string theory.\footnote{Note that the integral over q,q' will
reproduce the spacetime propagator for that mode of the string.}It
should be possible to reproduce the rest of the vertex operators after
imposing supersymmetry. 

\subsection{One point function for Wilson loop in a D-string background}
Let us now turn to the one point function of a general Wilson
loop operator in the presence of a D-string.  This was evaluated
also in \cite{IKKT}.  The D-string background can be written
as 
\[
A^{0}\; = \; ap_{1} + bq_{2}
\]
\[
A^{1} \; = \; cp_{2}+dq_{1}
\]
\be     \label{4.32} 
A^{I} \; = \; x^{I}
\ee
for some $a,b,c,d$ and $x^{I}$ is a constant matrix specifying
the location of the D-string which is in the $x^{1}$ direction.

We consider a Wilson loop representing a general closed string
configuration:\footnote{The continuum representation should
be equivalent to the `loop variables' introduced in \cite{BSLV}.}
$Tre^{KA}$.  
  The closed string world sheet coordinates being $q_{1}=\tau_{F}$
and $q_{2}=\sigma _{F}$, we can assume that $K^{\mu}$ is only a
function of $q_{2},p_{2}$.\footnote{This is analogous to the fact that
$e^{ikX(\tau )}$ is the vertex operator for a scalar particle
where the momentum $k$ does not depend on $\tau$.}
Thus 
\be     \label{4.33}
Tre^{KA} = \int dp_{1}dp_{2}dq_{1}dq_{2}e^{K^{0}(q_{2},p_{2})
(ap_{1} + bq_{2})
+K^{1}(q_{2},p_{2})(cp_{2} + dq_{1}) + K^{I}(q_{2},p_{2})x^{I}}
\ee
The $p_{1}$ and $q_{1}$ integrals clearly give $\delta (K^{0}(q_{2},p_{2}))
\delta (K^{1}(q_{2},p_{2}))$.  $K^{I}(q_{2},p_{2})$ contains, 
in addition to the
zero mode $K^{I}_{0}$, which just gives a factor
$e^{iK^{I}_{0}x^{I}}$, 
modes corresponding to the left and 
right movers. The integral over $q_{2}$ imposes the
$L_{0}=\bar{L_{0}}$ constraint.  Without knowing the precise form
of $K$ it is difficult to say more.  For instance if there are
necessarily factors of $p_{2}$ accompanying the non-zero modes, it
will
clearly give zero acting on $x^{I}$ which is proportional to the
identity.  

The delta functions in the 0 and 1 directions imply that
$\p _{\tau _{F}} X^{0}=\p _{\tau _{F}}X^{1}=0$ for the closed string.
By world sheet duality (s-channel - t-channel) this is equivalent to
$\p _{\sigma _{F}} X^{0} = \p _{\sigma _{F}}X^{1}=0$ for the open
string attached to the D-brane.  Similarly the factor
$e^{ik_{0}x^{I}}$ signifies the Dirichlet boundary conditions in the
other directions.  This calculation can thus be viewed as a 
matrix model derivation of the boundary conditions that define a
D-string in the conventional approach \cite{Polch}.

\newpage
\section{Conclusions}

In this paper we have attempted to explore some of the consequences of
the proposal of \cite{IKKT} identifying the reduced large N
(10 - dim) SSYM as a non-perturbative definition of type IIB superstring.
We showed that if we consider the Feynman diagrams of the quenched  
reduced model, a sensible genus expansion can be defined provided a double 
scaling limit is taken.  The small parameter is not $\frac{1}{N}$
but $\frac{1}{N \pld}$, where $\Lambda$ is proportional to the
volume of spacetime. A power of $m^{d}$ must be included for dimensional
reasons, $m$ being some mass parameter of the full theory.  The choice
$d=10$, reproduces the scaling proposed by \cite{IKKT}.  This
parameter can thus be identified with $g_{s}$ and the Wilson loop 
becomes the fundamental string.  We then introduced world sheet
coordinates $\sigma _{F}, \tau _{F}$ for the fundamental string
and argued that the D-string coordinates are essentially the
conjugates $\p _{\sigma _{F}},\p_{\tau _{F}}$ upto some overall
normalization.  The SL(2,Z) duality then becomes the usual symplectic
group of transformations acting on (p,q). The classical
D-string coresponds to turning on a background electric field.  By 
exchanging $p \leftrightarrow q $ and $g_{s} \leftrightarrow 
\frac{1}{g_{s}}$, one can also get a classical string solution.
The collective coordinate fluctuations can be quantized to get
the (m,n) strings.  We have not done this in a manifestly SL(2,Z) invariant
manner, but it should be possible to do so in this formalism.
Finally we have identified the Wilson loop equivalent of the
vertex operator $e^{ikX}$ and shown that the two point function
has the right form.  We have also calculated the one point function 
of a general closed string vertex operator in the presence of
a D-string.  This is tantamount to a derivation, within the
matrix model of the Dirichlet and Neumann boundary conditions that 
define D-branes in the usual approach.

There are many questions that need to be answered.  One principal 
issue is that of interactions.  This is to be derived from the
quartic term $[A^{I},A^{J}]^{2}$.  Since $[p_{i},q_{i}] \approx
\frac{1}{N_{i}}$ and $N_{1}N_{2}= N= \frac{1}{g_{s}}$, we
can naturally expect the coupling constant to show up here.
The details need to be worked out, to see whether it agrees with
the IIB superstring. 
This has been investigated in the context of the IIA string in \cite{Ver}.
 Finally, most of the formalism of this paper
can be applied to the IIA matrix model. This also needs to be worked out.

\noindent
{\bf Acknowledgements}\\
I would like to thank T. Jayaraman, G. Sengupta, N. Hari Dass and 
R. Kaul for useful discussions. I would also like to thank
T. Jayaraman
and N. Hari Dass for a careful reading of the manuscript and for
useful suggestions.

\end{document}